\documentclass[letterpaper, preprint, paper, 11pt]{AAS}	

\usepackage{bm}
\usepackage{amsmath, amsfonts, amssymb, mathtools}
\usepackage{subfigure}
\usepackage[colorlinks=true, pdfstartview=FitV, linkcolor=black, citecolor= black, urlcolor= black]{hyperref}
\usepackage{overcite}
\usepackage{footnpag}			      	

\usepackage{andrews_macro}

\PaperNumber{XX-XXX}

\begin{document}

\title{Optimal Fiducial Marker Placement for Satellite Proximity Operations Using Observability Gramians}

\author{Nicholas B. Andrews\thanks{Ph.D. Student, Department of Aeronautics and Astronautics, University of Washington, Seattle, WA 98195-2400} \ and  
Kristi A. Morgansen\thanks{Professor, Department of Aeronautics and Astronautics, University of Washington, Seattle, WA 98195-2400}}

\maketitle{} 		

\begin{abstract}
This paper investigates optimal fiducial marker placement on the surface of a satellite performing relative proximity operations with an observer satellite. The absolute and relative translation and attitude equations of motion for the satellite pair are modeled using dual quaternions. The observability of the relative dual quaternion system is analyzed using empirical observability Gramian methods. The optimal placement of a fiducial marker set, in which each marker gives simultaneous optical range and attitude measurements, is determined for the pair of satellites. A geostationary flyby between the observing body (chaser) and desired (target) satellites is numerically simulated and the optimal fiducial placement sets of five and ten on the surface of the desired satellite are solved. It is shown that the optimal solution maximizes the distance between fiducial markers and selects marker locations that are most sensitive to measuring changes in the state during the nonlinear trajectory, despite being visible for less time than other candidate marker locations. Definitions and properties of quaternions and dual quaternions, and parallels between the two, are presented alongside the relative motion model.
\end{abstract}

\section{Introduction}
With satellites becoming smaller and more autonomous than ever before, the role of the human operator and ground-based sensors in the loop is on the decline. The importance of accurate relative proximity operations between a pair(s) of satellites can already be seen when docking with the space station or during an inspection mission. Future space missions are investigating the use of teams of smaller, less expensive, and less capable satellites to replace the role of a single larger, more expensive, and more capable satellite. These proposed missions often require the team of satellites to orient themselves with respect to one another or a third-party satellite to accomplish tasks such as on-orbit repair, refueling, or surveillance \cite{Kristiansen2009}. During these missions, it is often difficult to resolve independent ground-based optical and range measurements for each satellite when they are sufficiently close together. Improving on-orbit sensing methodologies will improve relative state estimation accuracy and increase the overall efficiency and safety of the mission.

Modeling the pose (attitude and position) of a rigid body in a way that is efficient and tractable is applicable in many fields beyond satellite dynamics, such as robotics and computer vision \cite{Murray1994, Daniilidis1999}. It was shown that dual quaternions provide the most compact and computationally efficient representation for a rigid body transformation compared to other models \cite{Funda1990}. Dual quaternions can be thought of as an extension of unit quaternions, which are commonly used to represent attitude transformations. Computing coordinate transformations, relative states, and kinematic equations using dual quaternions have a similar form to quaternions and offer a convenient framework for working with rotating and/or moving reference frames \cite{Filipe2015, Stanfield2021}. Dual quaternions are also related to Chasles’s theorem which states that any pose transformation can be modeled as a \emph{screw motion}; a translation and rotation about a line \cite{Murray1994}.

The control and estimation of the dual quaternion satellite relative motion model have been investigated in previous works. Control of the coupled system with linear approximations was demonstrated by developing feedback linearization in \citenum{Wang2010} and applying a linear quadratic regulator (LQR) in \citenum{Stanfield2021}. An almost globally asymptotically stable nonlinear controller for tracking attitude and position was presented in \citenum{Filipe2013}, and later improved in \citenum{Filipe2015} when it was shown to perform well with uncertainty in satellite mass and inertia properties. Reference \citenum{Sabatini2011} proved observability via Lie derivatives of a nonlinear quaternion-only system using magnetic sensors. While \citenum{Zivan2022} showed observability of a linearized dual quaternion satellite proximity operation model with line-of-sight measurements.

The paper is organized as follows: first, mathematical preliminaries for quaternions and dual quaternions are presented. Fundamental operations, properties, and parallels between the two are discussed. The dual quaternion satellite relative motion model is then outlined in the following section. The measurement model for the fiducial markers is also introduced in this section. The Observability Tools section provides the background material necessary to understand what it means for a system to be observable, how to determine observability, as well as how to approximate the observability Gramian using empirical methods. The paper concludes by investigating optimal fiducial placement for a simulated geostationary flyby and with a discussion of the results.

\section{Mathematical Preliminaries}
An introduction to quaternion and dual quaternion math is presented below. There is a variety of notation and entry ordering used in quaternion and dual quaternion literature, so establishing the form used in this paper will be helpful in understanding the relative motion model discussed later. The material in this section can also be found in \cite{Filipe2015, Stanfield2021, Schaub2014} and a concise summary of definitions for basic quaternion and dual quaternion operations is shown in table \ref{tab:quaternion}.
\begin{table}[ht]
    \centering
    \makebox[\textwidth][c]{
    \begin{tabular}{|c | c | c|}
        \hline
        \textbf{Operation} & \textbf{Quaternion Definition} & \textbf{Dual Quaternion Definition} \\
        \hline
        Addition & $\qblank{a} + \qblank{b} = (\qscalar{a} + \qscalar{b}, \ \qvec{a} + \qvec{b})$ & $\dualqblank{a} + \dualqblank{b} = (\qreal{a} + \qreal{b}) + \dualunit (\qdual{a} + \qdual{b})$ \\
        Scalar Multiplication & $\lambda \qblank{a} = (\lambda \qscalar{a}, \ \lambda \qvec{a})$ & $\lambda \dualqblank{a} = (\lambda \qreal{a}) + \dualunit (\lambda \qdual{a})$ \\
        Multiplication & $\qblank{a} \qblank{b} = (\qscalar{a} \qscalar{b} - \qvec{a} \qdot \qvec{b}, \ \qscalar{a} \qvec{b} + \qscalar{b} \qvec{a} + \qvec{a} \cross \qvec{b})$ & $\dualqblank{a} \dualqblank{b} = (\qreal{a} \qreal{b}) + \dualunit (\qdual{a} \qreal{b} + \qreal{a} \qdual{b})$ \\
        Conjugate & $\qconj{\qblank{a}} = (\qscalar{a}, \ -\qvec{a})$ & $\qconj{\dualqblank{a}} = (\qconj{\qreal{a}}) + \dualunit (\qconj{\qdual{a}})$ \\
        Dot Product & $\qblank{a} \qdot \qblank{b} = (\qscalar{a} \qscalar{b} + \qvec{a} \qdot \qvec{b}, \ \zeros{3}{1} )$ & $\dualqblank{a} \qdot \dualqblank{b} = (\qreal{a} \qdot \qreal{b}) + \dualunit (\qdual{a} \qdot \qreal{b} + \qreal{a} \qdot \qdual{b})$ \\
        Cross Product & $\qblank{a} \cross \qblank{b} = (0, \ \qscalar{a} \qvec{b} + \qscalar{b} \qvec{a} + \qvec{a} \cross \qvec{b})$ & $\dualqblank{a} \cross \dualqblank{b} = (\qreal{a} \cross \qreal{b}) + \dualunit (\qdual{a} \cross \qreal{b} + \qreal{a} \cross \qdual{b})$ \\
        Norm & $\norm{\qblank{a}} = \sqrt{\qblank{a} \qdot \qblank{a}}$ & $\norm{\dualqblank{a}} = \sqrt{(\qreal{a} \qdot \qreal{a} + \qdual{a} \qdot \qdual{a}) + \dualunit 0}$ \\
        Swap & Undefined & $\swap{\dualqblank{a}} = \qdual{a} + \dualunit \qreal{a}$ \\
        \hline
    \end{tabular}
    }
    \caption{Quaternion and dual quaternion operations \cite{Stanfield2021}.}
    \label{tab:quaternion}
\end{table}

\subsection{Quaternions}
A quaternion is defined as:
\begin{align}
    \qblank{q} = \qscalar{q} + \qblank{q_1} i + \qblank{q_2} j + \qblank{q_3} k
\end{align}
Where $i^2 = j^2 = k^2 = -1$. A quaternion can be represented as its scalar part $\qscalar{q} \in \reals{}$ and vector part $\qvec{q} =  \left[ \qblank{q_1}, \ \qblank{q_2}, \ \qblank{q_3} \right]^\top \in \reals{3}$, succinctly expressed as a stacked vector $\qblank{q} = (\qscalar{q}, \ \qvec{q}) \in \reals{4}$. In some literature, the scalar component will be placed at the end of the quaternion vector. However, for this paper, the scalar component will always be assumed to be the first entry.

The relative orientation of a frame $\cf{x}$ with respect to the frame $\cf{y}$ can be represented as the \emph{unit quaternion} (or rotation quaternion) $\q{x}{y}$. The primary advantage of unit quaternions compared to Euler angles or other rotation representations is that quaternions are singularity free. A unit quaternion describing the relative orientation of $\cf{x}$ relative to $\cf{y}$ is defined in terms of a rotation angle $\phi$ about the unit vector $\ivec{n}$:
\begin{align}
    \q{x}{y} = \left( \cos \left( \frac{\phi}{2} \right), \ \ivec{n} \sin \left( \frac{\phi}{2} \right) \right)
\end{align}
A unit quaternion satisfies the following properties:
\begin{equation}
    \label{eq:unitq}
    \begin{gathered}
        \qconj{\q{x}{y}} \q{x}{y} = \q{x}{y} \qconj{\q{x}{y}} = \qones \\
        \inv{\q{x}{y}} = \qconj{\q{x}{y}} = \q{y}{x}
    \end{gathered}
\end{equation}
where $\qones = (1, \ \zeros{3}{1})$. Similar to how the inverse is the transpose for rotation matrices, the inverse is the conjugate for unit quaternions.

Unit quaternions have a convenient form for changing vector reference frames. Redefining a vector $\qvecf{u}{x} \in \reals{3}$ in the $\cf{x}$ frame as $\qblankf{u}{x} = (0, \ \qvecf{u}{x}) \in \reals{4}$, then a coordinate transformation to and from the $\cf{y}$ frame has the form of:
\begin{equation}
    \label{eq:quatrot}
    \begin{gathered}
        \qblankf{u}{y} = \qconj{\q{y}{x}} \qblankf{u}{x} \q{y}{x} \\
        \qblankf{u}{x} = \q{y}{x} \qblankf{u}{y} \qconj{\q{y}{x}}
    \end{gathered}
\end{equation}
Additionally, unit quaternions can be chained together to solve for the total relative rotation between multiple reference frames:
\begin{align}
    \label{eq:qchain}
    \q{x}{y} = \qconj{\q{y}{z}} \q{x}{z}
\end{align}

The unit quaternion kinematic equations for a rotating frame are
\begin{align}
    \label{eq:qkin}
    \dq{x}{y} = \frac{1}{2} \q{x}{y} \qomega{x}{y}{x} = \frac{1}{2} \qomega{x}{y}{y} \q{x}{y}
\end{align}
where $\omeg{x}{y}{x} \in \reals{3}$ is the angular velocity of $\cf{x}$ relative to $\cf{y}$ in $\cf{x}$ coordinates and $\qomega{x}{y}{x} = (0, \ \omeg{x}{y}{x}) \in \reals{4}$.

\subsection{Dual Quaternions}
Dual quaternions are an extension of quaternions and provide a convenient and natural form to model the relative pose (orientation and translation), and their respective velocities, through a compact coupling of the rotational and translational kinematics. Similar to how a complex number is composed of a real and imaginary part, a dual quaternion is formed from a real and dual part. The dual unit $\dualunit$ signifies the dual part of the dual quaternion. A dual quaternion $\dualqblank{q} \in \reals{8}$ is defined as
\begin{align}
    \dualqblank{q} = \qreal{q} + \dualunit \qdual{q}
\end{align}
where $\qreal{q} \in \reals{4}$ and $\qdual{q} \in \reals{4}$ are quaternions representing the real and dual components. Like the unit quaternion properties in equation \ref{eq:unitq}, a \emph{unit dual quaternion} satisfies similar unit length and inverse properties shown below where $\dualones = \qones + \dualunit \qzeros$.
\begin{equation}
    \begin{gathered}
        \qconj{\dualq{x}{y}} \dualq{x}{y} = \dualq{x}{y} \qconj{\dualq{x}{y}} = \dualones \\
        \inv{\dualq{x}{y}} = \qconj{\dualq{x}{y}} = \dualq{y}{x}
    \end{gathered}
\end{equation}

Similar to the quaternion form in equation \ref{eq:quatrot}, changing reference frames for a dual quaternion $\dualo{x}{y}{x} \in \reals{8}$ can be done as follows:
\begin{equation}
    \begin{gathered}
        \dualo{x}{y}{y} = \qconj{\dualq{y}{x}} \dualo{x}{y}{x} \dualq{y}{x} \\
        \dualo{x}{y}{x} = \dualq{y}{x} \dualo{x}{y}{y} \qconj{\dualq{y}{x}}
    \end{gathered}
\end{equation}
Like unit quaternions in equation \ref{eq:qchain}, unit dual quaternions can also be chained together to solve for a total transformation between intermediate frames.
\begin{align}
    \dualq{x}{y} = \qconj{\dualq{y}{z}} \dualq{x}{z}
\end{align}

The \emph{dual position} is a unit dual quaternion that describes the relative pose between coordinate frames, it is defined as
\begin{align}
    \dualq{x}{y} = \q{x}{y} + \dualunit \frac{1}{2} \qpos{x}{y}{y} \q{x}{y} = \q{x}{y} + \dualunit \frac{1}{2} \q{x}{y} \qpos{x}{y}{x} 
\end{align}
where $\qpos{x}{y}{x} = (0, \ \pos{x}{y}{x}) \in \reals{4}$ and $\pos{x}{y}{x} \in \reals{3}$ is the position of $\cf{x}$ relative to $\cf{y}$ expressed in $\cf{x}$ coordinates.

The \emph{dual velocity} is a dual quaternion and describes the relative rotational and translational velocities. It is derived from the transport theorem and has the general form
\begin{align}
    \dualo{x}{y}{z} = \qomega{x}{y}{z} + \dualunit (\qvel{x}{y}{z} + \qomega{x}{y}{z} \cross \qpos{z}{x}{z})
\end{align}
where $\qvel{x}{y}{z} = (0, \ \vel{x}{y}{z}) \in \reals{4}$ and $\vel{x}{y}{z} \in \reals{3}$ is the translational velocity of $\cf{x}$ relative to $\cf{y}$ expressed in $\cf{z}$ coordinates. Unlike unit dual quaternions, calculating relative dual quaternions are treated similarly to vectors and have the form
\begin{align}
    \dualo{x}{y}{z} = \dualo{x}{w}{z} - \dualo{y}{w}{z}
\end{align}

The dual quaternion kinematics in equation \ref{eq:dqkin} also have a nice parallel to the quaternion kinematics in equation \ref{eq:qkin}. Despite the familiar form of rotation-only kinematics, it is important to remember that the dual quaternion kinematics reflect the rotation and translation because of the coupling in the dual quaternion formulation.
\begin{align}
    \label{eq:dqkin}
    \ddualq{x}{y} = \frac{1}{2} \dualq{x}{y} \dualo{x}{y}{x} = \frac{1}{2} \dualo{x}{y}{y} \dualq{x}{y}
\end{align}

Lastly, the multiplication of a block matrix $M \in \reals{8 \times 8}$ with a dual quaternion is defined as
\begin{equation}
    \begin{gathered}
        M = \mtx{M_{11} & M_{12} \\ M_{21} & M_{22}}, \quad M_{11}, \ M_{12}, \ M_{21}, \ M_{22} \in \reals{4 \times 4} \\
        M \dualstar \dualqblank{q} = (M_{11} * \qreal{q} + M_{12} * \qdual{q}) + \dualunit (M_{21} * \qreal{q} + M_{22} * \qdual{q})
    \end{gathered}
\end{equation}
This operation is equivalent to right multiplying a $M$ by a $\reals{8}$ vector.

\section{Relative Motion Model} \label{eq:model}
\begin{figure}[ht]
    \centering
    \includegraphics[scale=0.5]{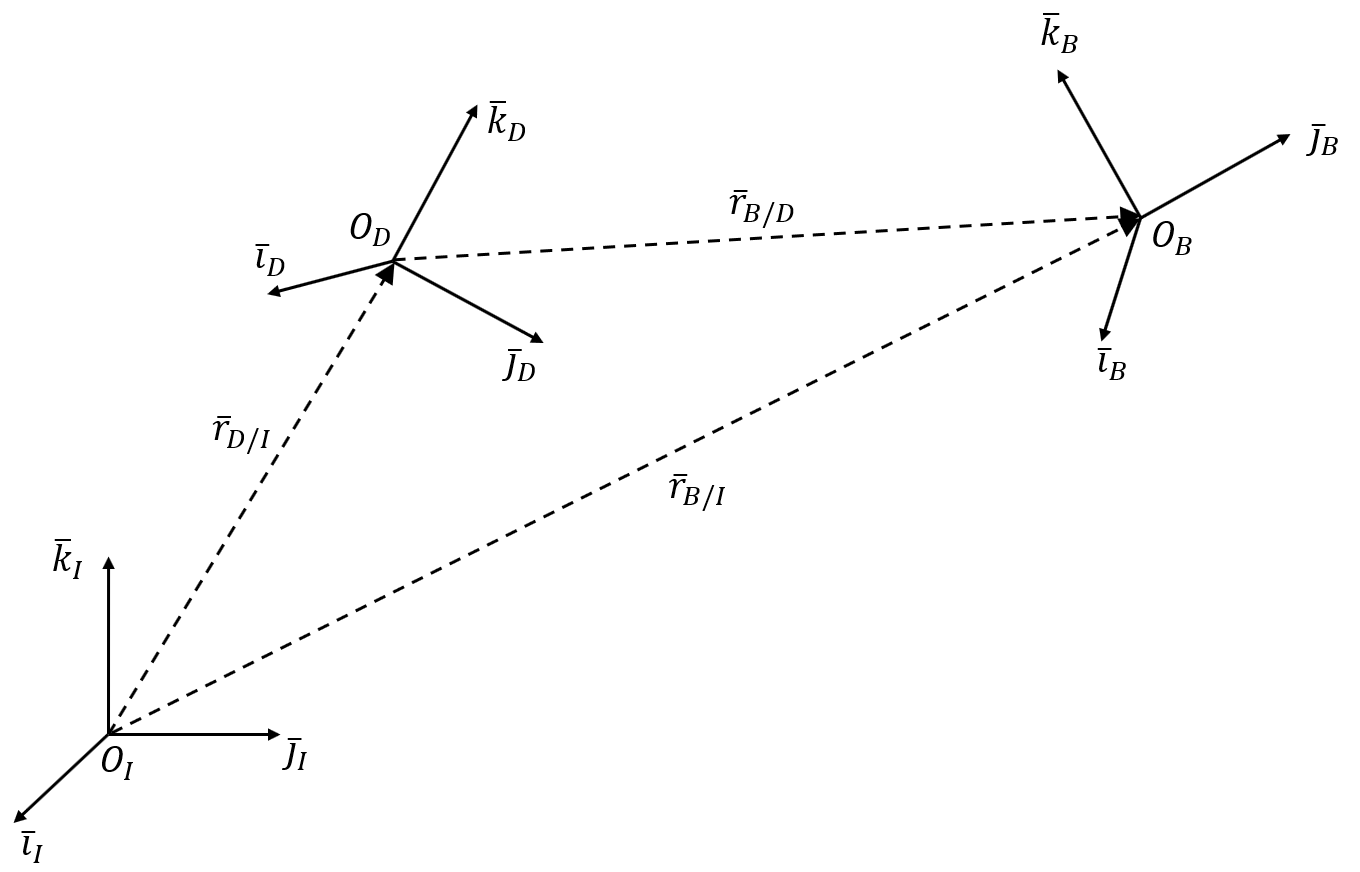}
    \caption{Model of relative spacecraft system.}
    \label{fig:rpo_model}
\end{figure}

For satellite relative proximity operations we are more concerned with tracking the relative pose and velocities between pairs of satellites, and less with their state with respect to an inertial frame. There are three coordinate frames used in the relative motion problem between a pair of satellites: body $\body$, desired $\desired$, and inertial $\inertial$. The body frame is fixed to the chaser (secondary) spacecraft and the desired frame is fixed to the target (primary) spacecraft. The states of interest are the relative pose and velocities of the body with respect to the desired. It is assumed in this paper that both spacecraft are in orbit around the Earth and the inertial frame is the J2000 Earth-centered inertial (ECI) frame. A depiction of the relative motion problem outlined above is shown in figure \ref{fig:rpo_model}.

Perhaps the most straightforward method for modeling this system is to treat the translational and rotational equations of motion separately for the body and desired, solve for their states with respect to the inertial frame, and then compute the desired relative state while accounting for the moving/rotating reference frames of the body and desired. However, this is a roundabout way to our desired state representation and requires tedious and sometimes intractable calculations between coordinate frames. An alternative is to use the Clohessy-Wiltshire or Hill equations to approximate the nonlinear relative translational motion with a linear model \cite{Schaub2014, Vallado2013}. However, this approach assumes the body and desired frames are relatively close together, the desired is in a circular orbit, the body is in a circular or elliptical orbit, and the translational and rotational equations of motion are treated separately.

The dual quaternion approach improves the efficiency and accuracy of the relative motion problem by providing a convenient framework for computing relative states and simulating the relative equations of motion. One of the most powerful features of embedding the pose and velocities in the dual position and dual velocity formats is that velocities between fixed, moving, or rotating frames can be calculated at a "higher" level through dual quaternion operations, and without the need to ever explicitly compute them via the transport theorem. However, a reference trajectory of the desired with respect to inertial is still required for computing external forces, but the full nonlinear dynamics can be modeled without simplifying assumptions. The pose and velocities of the desired with respect to inertial are assumed to be known. This assumption is motivated by the reality that the state of the desired spacecraft is often well-characterized (e.g. geostationary communication satellite).

\subsection{Equations of Motion}
The equations of motion for the dual quaternion relative motion model are \cite{Filipe2015, Stanfield2021}:
\begin{align}
    \ddualq{b}{d} &= \frac{1}{2} \dualq{b}{d} \dualo{b}{d}{b} \\
    \swap{(\ddualo{b}{d}{b})} &= \inv{(\mass)} \dualstar (\dualforce{} - (\dualo{b}{d}{b} + \dualo{d}{i}{b}) \cross (\mass \dualstar (\swap{(\dualo{b}{d}{b})} + \swap{(\dualo{d}{i}{b})})) \\
    &- \mass \swap{(\qconj{\dualq{b}{d}} \ddualo{d}{i}{d} \dualq{b}{d})} - \mass \dualstar \swap{(\dualo{d}{i}{b} \cross \dualo{b}{d}{b})}) \nonumber
\end{align}
The mass matrix $\mass$ is defined in equation \ref{eq:mass} where $m$ is the mass of the body and $\inertia \in \reals{3 \times 3}$ is the inertia matrix of the body in $\cf{b}$ frame coordinates.
\begin{align}
    \label{eq:mass}
    \mass = \mtx{
    1 & \zeros{1}{3} & 0 & \zeros{1}{3} \\
    \zeros{3}{1} & m \eye{3} & \zeros{3}{1} & \zeros{3}{3} \\
    0 & \zeros{1}{3} & 1 & \zeros{1}{3} \\
    \zeros{3}{1} & \zeros{3}{3} & \zeros{3}{1} & \inertia
    }
\end{align}

The dual force $\dualforce{}$ represents all external forces $\qforce{}$ and torques $\qtorque{}$ acting on the relative motion system. For on-orbit relative proximity operations, the primary dual forces that could be acting on the system are gravitational acceleration $\dualforce{g}$, $\jtwo$ perturbing forces due to Earth's oblateness $\dualforce{\jtwo}$, gravity gradient torque $\dualforce{\gradg}$, and control inputs $\dualforce{c}$. The total dual force is simply the sum of the individual dual forces:
\begin{equation}
    \begin{gathered}
        \dualforce{} = \qforce{} + \dualunit \qtorque{} \\
        \qforce{} = (0, \ \force{}), \quad \qtorque{} = (0, \ \torque{}) \\
        \dualforce{} = \dualforce{g} + \dualforce{\jtwo} + \dualforce{\gradg} + \dualforce{c}
    \end{gathered}
\end{equation}

The dual force due to gravitational acceleration is:
\begin{equation}
    \begin{gathered}
        \dualforce{g} = m \dualaccbody{g} = \mass \dualstar \dualaccbody{g} \\
        \dualaccbody{g} = \qaccbody{g} + \dualunit \qzeros \\
        \qaccbody{g} = (0, \ \acc{g}) \\
        \acc{g} = -\mu \frac{\pos{b}{i}{b}}{\norm{\pos{b}{i}{b}}^3}
    \end{gathered}
\end{equation}
Where $\mu$ is the gravitational parameter of the central body. The dual force due to $\jtwo$ is:
\begin{equation}
    \begin{gathered}
        \dualforce{\jtwo} = m \dualaccbody{\jtwo} = \mass \dualstar \dualaccbody{\jtwo} \\
        \dualaccbody{\jtwo} = \qaccbody{\jtwo} + \dualunit \qzeros \\
        \qaccbody{\jtwo} = (0, \ \acc{\jtwo}) \\
        \acc{\jtwo} = -\frac{3 \mu \jtwo \re^2}{2 \norm{\pos{b}{i}{b}}^5}
        \mtx{
        \left( 1 - 5 \left( \frac{\escal{z}{b}{i}{b}}{\norm{\pos{b}{i}{b}}} \right)^2 \right) \escal{x}{b}{i}{b} \\
        \left( 1 - 5 \left( \frac{\escal{z}{b}{i}{b}}{\norm{\pos{b}{i}{b}}} \right)^2 \right) \escal{y}{b}{i}{b} \\
        \left( 3 - 5 \left( \frac{\escal{z}{b}{i}{b}}{\norm{\pos{b}{i}{b}}} \right)^2 \right) \escal{z}{b}{i}{b}
        }
    \end{gathered}
\end{equation}
Where $\pos{b}{i}{b} = \tpose{\mtx{\escal{x}{b}{i}{b} & \escal{y}{b}{i}{b} & \escal{z}{b}{i}{b}}}$. The dual gravity gradient torque is:
\begin{equation}
    \begin{gathered}
        \dualforce{\gradg} = \qzeros + \dualunit \qtorque{\gradg} = \frac{3 \mu \edual{r}{b}{i}{b}}{\norm{\edual{r}{b}{i}{b}}^5} \cross \left( \mass \dualstar \swap{(\edual{r}{b}{i}{b})} \right) \\
        \qtorque{\gradg} = (0, \ \torque{\gradg}) \\
        \torque{\gradg} = 3 \mu \frac{\pos{b}{i}{b} \cross (\inertia \pos{b}{i}{b})}{\norm{\pos{b}{i}{b}}^5}
    \end{gathered}
\end{equation}
Where $\edual{r}{b}{i}{b} = \qpos{b}{i}{b} + \dualunit \qzeros$. Lastly, the dual control input takes the same form as the dual forces and torques, but the control law is left to the operator's discretion:
\begin{equation}
    \begin{gathered}
        \dualforce{c} = \qforce{c} + \dualunit \qtorque{c} \\
        \qforce{} = (0, \ \force{c}), \quad \qtorque{} = (0, \ \torque{c})
    \end{gathered}
\end{equation}
The model used in this paper considers gravitational acceleration and $\jtwo$ perturbing forces due to Earth's oblateness as the only external forces and is rewritten for completeness as
\begin{align}
    \dualforce{} = \dualforce{g} + \dualforce{\jtwo}
\end{align}

\subsection{Measurement Model}
\begin{figure}[ht]
    \centering
    \includegraphics[scale=0.5]{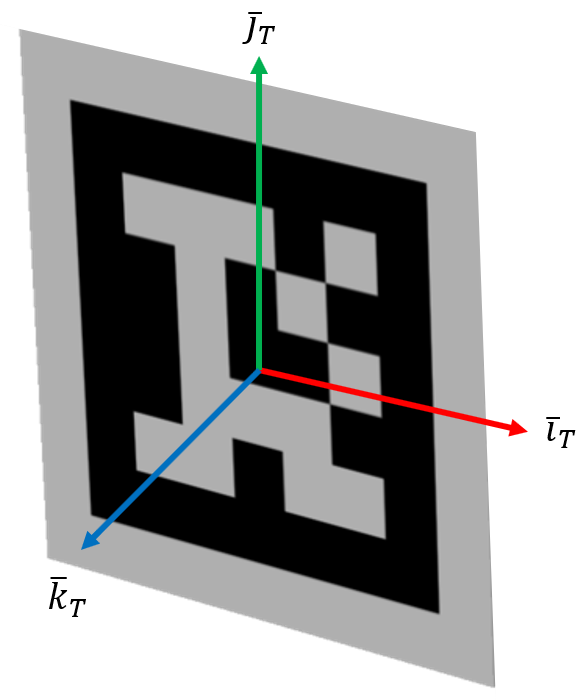}
    \caption{AprilTag fiducial marker and tag coordinate frame $\cf{t}$ \cite{Honigmann2020}.}
    \label{fig:tagframe}
\end{figure}

The sensor methodology investigated in this research is the use of fiducial markers for relative proximity operations. The fiducial markers are placed on the desired spacecraft and observed by the body spacecraft. This sensor methodology is motivated by AprilTag fiducial markers \cite{Wang2016} and a sample tag with its respective coordinate frame $\cf{t}$ is shown in figure \ref{fig:tagframe}. When placed in front of a camera, a single fiducial marker gives a range and attitude measurement. The range is measured with respect to the center of the tag's face and the attitude is represented as a relative unit quaternion. Both the range $\range$ and attitude $\q{b}{t}$ measurements are observed from the body frame $\cf{b}$ and measured with respect to the tag frame $\cf{t}$, which is affixed to the surface of the desired satellite and displaced from the origin of the desired frame $\cf{d}$. The measurement model for a single tag has the nonlinear form:
\begin{equation}
    \begin{aligned}
        y &= h(x) \\
        \mtx{\range \\ \q{b}{t}} &= h \left( \mtx{\dualq{b}{d} \\ \dualo{b}{d}{b}} \right) \in \reals{5} \\
        &= \mtx{\norm{\pos{b}{t}{t}} \\ \qconj{\q{t}{d}} \q{b}{d}} \\
        &= \mtx{\norm{\qpos{b}{d}{b} - \qconj{\q{b}{d}} \qpos{t}{d}{d} \q{b}{d}} \\ \qconj{\q{t}{d}} \q{b}{d}}
    \end{aligned}
\end{equation}

A body with multiple tags will return a set of measurement vectors with cardinality equal to the number of visible tags from the observer's perspective. For a tag to be considered visible it must meet a minimum elevation constraint. The elevation angle $\phi$ is measured off the face of the tag and is defined as:
\begin{align}
    \phi = \arcsin \left( \frac{\pos{b}{t}{t} \cdot \tpose{\mtx{0 & 0 & 1}}}{\norm{\pos{b}{t}{t}}} \right)
\end{align}

\section{Observability Tools} \label{sec:observability}
Observability describes the feasibility of uniquely determining an initial state of a system from measurements over a finite time interval. If the unknown initial state $x_0 \in \reals{n}$ can be uniquely determined in an open neighborhood of $x_0$ from the outputs $y \in \reals{m}$, then the system is \emph{weakly observable} \cite{Powel2015}. The observability of linear systems can be evaluated analytically using the observability matrix and Gramian, whereas a computational approach can be taken to calculate the empirical observability Gramian for nonlinear and analytically intractable systems. Numerical observability tools provide measures of observability that can be used to determine whether or not a system is observable and assess performance for optimal sensor placement.

\subsection{Linear Observability}
In linear systems (equation \ref{eq:linsys}), the observability matrix, $\obsv$, is obtained by differentiating the output $y$ in a linear system with respect to time and collecting terms that are multiplied by the state $x$:
\begin{equation}
    \label{eq:linsys}
    \begin{gathered}
        \dot{x} = Ax + Bu  \\
        y = Cx + Du
    \end{gathered} 
\end{equation}
\begin{equation}
    \obsv =  \mtx{C \\ CA \\ \vdots \\ CA^{n-1}}
\end{equation}
A linear system is then observable if and only if $\obsv$ is full rank \cite{Hespanha2009, Astrom2020}.

Another tool to determine observability is the observability Gramian, which quantitatively captures the sensitivity of the measurements to changes in the initial conditions. The standard expression of the observability Gramian for a linear time-invariant (LTI) system is 
\begin{align}
    \og(t) &= \int_{0}^{t} e^{A^\top \tau}C^\top C e^{A \tau} \ d\tau \label{eq:Wlinear}
\end{align}
If $\og(t) \in \reals{n \cross n}$ is nonsingular for some $t>0$, the system will be observable. Furthermore, the eigenvector associated with the largest (smallest) eigenvalue indicates the mode that is most (least) observable \cite{Krener2009, Hinson2014, Powel2015}.

The observability Gramian can be re-written in terms of measurements by using the solution $y(t)$ to the linear system (\ref{eq:linsys}):
\begin{gather}
    y(t) = C e^{At} x_0 + \int_{0}^{t} C e^{A(t - \tau)} B u(\tau) \ d\tau + D u(t) \label{eq:ysol}
\end{gather}
where $x_0 = x(0)$ \cite{Hespanha2009, Astrom2020}. Then differentiating \ref{eq:ysol} with respect to $x_0$ and substituting in to \ref{eq:Wlinear} yields:
\begin{gather}
    \og(t) = \int_{0}^{t} \tpose{\pd{y(\tau)}{x_0}} \pd{y(\tau)}{x_0} \ d\tau \label{eq:Wlineary}
\end{gather}
The partial derivative $\pd{y(t)}{x_0} \in \reals{m \times n}$ is a Jacobian matrix where it's $i$th column is the derivative of the measurement vector with respect to the $i$th entry in $x_0$.
\begin{align}
    \pd{y(t)}{x_0} = \mtx{\pd{y(t)}{x_1(0)} & \pd{y(t)}{x_2(0)} & \cdots & \pd{y(t)}{x_n(0)}}
\end{align}

\subsection{Empirical Observability Gramian}
The empirical observability Gramian provides a method to approximate $\og(t)$ for nonlinear systems or for linear systems where the Gramian is difficult to calculate. The sensitivity of the measurements to changes in the initial condition (\ref{eq:Wlineary}) is approximated by using the central difference method and individually perturbing each entry in the initial state by a small value $\epsilon$. The empirical observability Gramian is \cite{Krener2009, Hinson2014, Powel2015}
\begin{equation}
    \eog(t) =\frac{1}{4\epsilon^2} \sum_{\tau = 0}^{t} 
    \Delta Y(\tau)^\top \Delta Y (\tau)
    \label{eqn:empGram}
\end{equation}
where $\Delta Y (t) = \mtx{\Delta y^{\pm1}(t) & \Delta y^{\pm2}(t) & \cdots & \Delta y^{\pm n}(t) } \in \reals{m \times n}$ is comprised of the differences in the outputs, $\Delta y^{\pm i}(t) = y^{+i}(t) - y^{-i}(t)$,  that result from perturbing the $i$th row in the initial condition $x_0$ by $\pm \epsilon \mathbf{e}_i$. Where $\mathbf{e}_i$ is a zero vector with $1$ in the $i$th row. If the empirical observability Gramian, $\og^\epsilon(t) \in \mathbb{R}^{n \times n}$, is full rank at the limit $\epsilon\to0$, then the system is weakly observable at $x_0$ \cite{Powel2015}.

\subsection{Measures of Observability}
The following metrics quantify the degree of observability based on the observability Gramian and allow for optimization over a potential sensor set by casting the metric as a cost function in a minimization problem \cite{Krener2009}.

\subsubsection{Minimum Eigenvalue}
This measure is the reciprocal of the minimum eigenvalue and provides a measure of the least observable mode
\begin{align}
    J_{\nu}(\og) = \frac{1}{\lambda_{min}(\og)} = \nu.
\end{align}
Minimizing this cost maximizes the observability of the least observable mode will be.

\subsubsection{Condition Number}
This measure is the ratio between the largest and smallest eigenvalues
\begin{align}
    J_{\kappa}(\og) = \frac{\lambda_{max}(\og)}{\lambda_{min}(\og)} = \kappa.
\end{align}
The closer to one this measure is, the more balanced the information from the sensor is and the better conditioned the inversion of the map from states to measurements will be. The condition number is not a viable metric on its own and must be used with some caution as it may prioritize minimizing $\lambda_{max}$.

\subsection{Optimal Sensor Placement}
With binary sensor activation variables $\alpha_i \in \{0,1\}$ and $\alpha \in \reals{p}$ that indicate if a sensor is in use, define the total observability Gramian as the sum $\ogtilde(\alpha) = \sum_{i=1}^{p} W_i^\epsilon \alpha_i$ where $p$ is the number of potential sensor locations and $W_i^\epsilon = \eog (t, i)$ is calculated according to equation \ref{eqn:empGram}, but only using the $i$th entry in the measurement vector $y$. This decomposition results in a single $\reals{n \times n}$ empirical observability Gramian for each candidate sensor location and makes it possible to optimize over different combinations of sensors. The optimal sensor placement problem for determining the optimal subset of $c$ sensors from a set of $p$ feasible sensors can be thought of as choosing the best combination of $c$ observability Gramians. The optimization problem we wish to solve is: \cite{Hinson2014, Brace2022}
\begin{eqnarray}
 \min_{\alpha} & & J(\ogtilde (\alpha) ) \nonumber  \\
\text{subject to}  & & 0 = \tpose{1} \alpha - c \label{eq:optprob} \\
                   & & \alpha_{i} \in \{0,1\}  \nonumber
\end{eqnarray} 

This formulation is, however, a non-convex mixed integer program, making it difficult to solve and without global optimality guarantees. By easing the requirement that the sensor activation variable, $\alpha_i \in \{0,1\}$, be binary and instead requiring a value $0 \leq a_i \leq 1$, the constraints become convex. If the objective function is convex with respect to the activation variables, the optimization problem is then convex. The measures discussed in the previous section are convex ($\nu$) or quasiconvex ($\kappa$) with respect to the variables $a_i$ \cite{Boyd2004}. This relaxation from binary to continuous sensor activation variables makes the solution tractable, however, it is generally sub-optimal.

For this paper, the objective function considered is maximizing the minimum eigenvalue of the observability Gramian. After relaxing the activation variables and applying the identity $\lambda_{max}(\ogtilde(a)) \eye{n} \succeq \ogtilde(a) \succeq \lambda_{min}(\ogtilde(a)) \eye{n}$, the optimization problem becomes the semidefinite program shown in equation \ref{eq:mineig} \cite{Boyd2004}. Once the solution is found, the optimal set of sensors is chosen to be those corresponding to the largest $c$ values in the activation vector $a$.
\begin{eqnarray}
\label{eq:mineig}
 \max_{a, \lambda} & & \lambda  \nonumber\\
\text{subject to}  & & \ogtilde(a) - \lambda \eye{n} \succeq 0 \label{eq:optcvx} \\
                   & & 0 \leq a \leq 1 \nonumber \\
                   & & 0 = \tpose{1} a - c \nonumber
\end{eqnarray}

\section{Simulation Results}
\subsection{Problem Setup}
\begin{figure}[ht]
    \centering
    \includegraphics[height=3.5in]{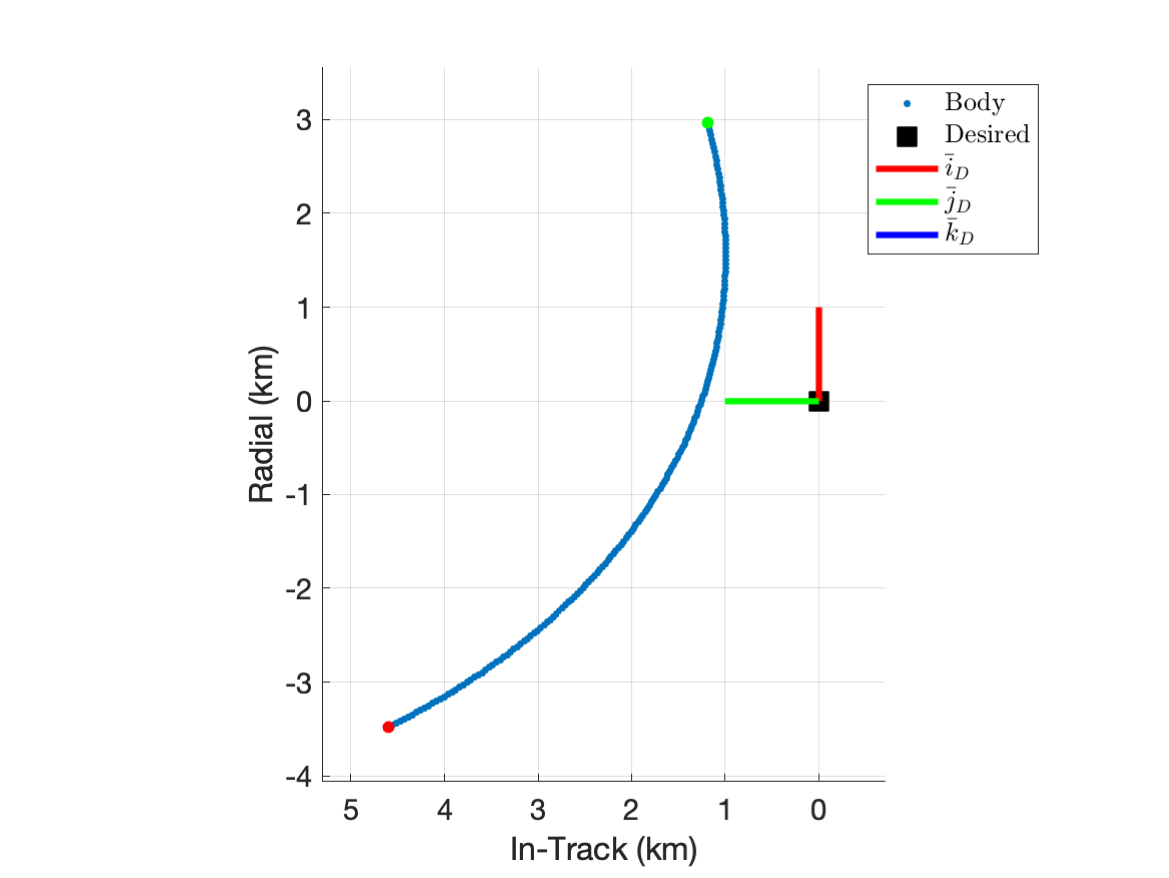}
    \caption{Nominal flyby trajectory from +R/+I to -R/+I.}
    \label{fig:ric}
\end{figure}

The simulated scenario is a geostationary flyby sampled once per minute over a 3-hour window. The desired frame $\cf{d}$ is aligned with the Radial, In-track, Cross-track (RIC) frame $\cf{r}$ to simulate a nadir-pointing spacecraft \cite{Vallado2013}. For the sake of simplicity and since there are no visibility constraints in play that are a function of the body's attitude, the body frame remains aligned with the ECI frame for the duration of the simulation. A plot of the flyby trajectory in the RIC frame centered about the desired spacecraft is shown in figure \ref{fig:ric}. The colored vectors centered on the desired spacecraft represent the principal axis of the desired frame $\cf{d}$, which will be useful later as a reference to comment on the optimal fiducial marker placement results. The colored points at the ends of the flyby trajectory indicate the start (green) and stop (red) points over the 3-hour simulation interval. The initial conditions for the body and desired spacecraft are defined with respect to the inertial frame and are shown in table \ref{tab:init}.
\begin{table}[ht]
    \centering
    \makebox[\textwidth][c]{
    \begin{tabular}{|c | c | c|}
        \hline
        \textbf{Parameter} & \textbf{Body $\cf{b}$} & \textbf{Desired $\cf{d}$} \\
        \hline
        $\pos{\cdot}{i}{i}$ \ (km) & $\tpose{\mtx{-17517.33 &
-38359.24 & 0}}$ & $\tpose{\mtx{-17517.18 & -38356.04 & 0}}$ \\
        $\vel{\cdot}{i}{i}$ \ (km/s) & $\tpose{\mtx{2.80 & -1.28 & 0}}$ & $\tpose{\mtx{2.80 & -1.28 & 0}}$ \\
        $\q{\cdot}{i}$ & $\qones$ & $\tpose{\mtx{0.54 & 0 & 0 & -0.84}}$ \\
        $\omeg{\cdot}{i}{i}$ \ (rad/s) & $\zeros{3}{1}$ & $\tpose{\mtx{0 & 0 & 7.29 \mathrm{e}{-5}}}$ \\
        $m$ \ (kg) & 10 & Undefined \\
        $\inertia$ \ ($\text{kg m}^2$) & $\eye{3}$ & Undefined \\
        \hline
    \end{tabular}
    }
    \caption{Initial conditions.}
    \label{tab:init}
\end{table}

\begin{figure}[ht]
    \centering
    \includegraphics[height=3in]{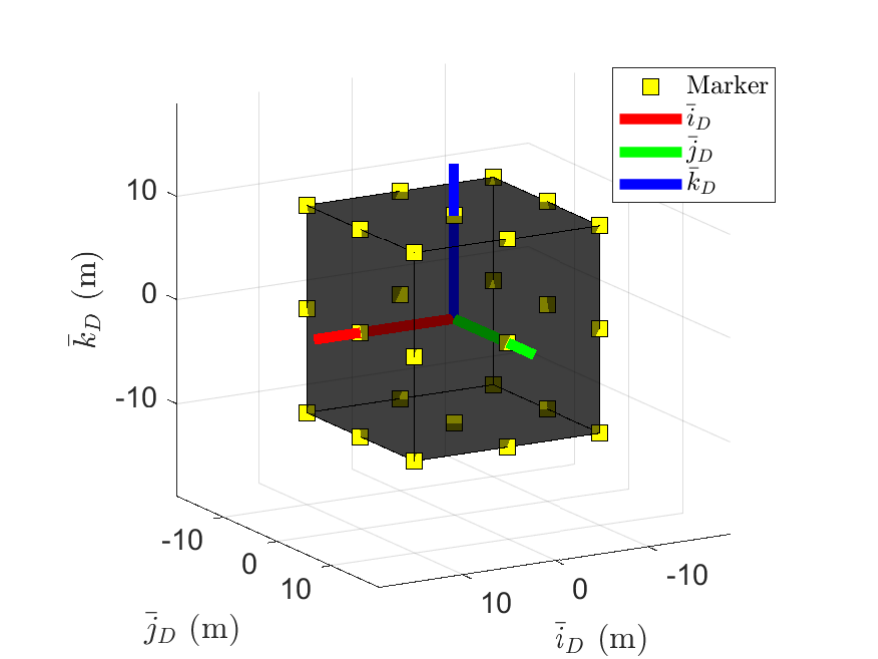}
    \caption{Feasible fiducial marker set on the desired spacecraft.}
    \label{fig:alltag}
\end{figure}

The desired spacecraft is modeled as a cube with 20m sides. Candidate fiducial marker locations are equally spaced across each face of the cube at 10m intervals, for a total of 9 tags per face and 64 tags in total. A minimum elevation angle of $30 \degree$ off the face of the tag is the only visibility constraint for a valid measurement. The feasible set of tag locations is shown in figure \ref{fig:alltag}. Once again, the colored vectors extending from the center of the cube are the principal axis of the desired frame $\cf{D}$.

The initial conditions are first transformed into dual positions and velocities with respect to the inertial frame:
\begin{equation}
    \begin{gathered}
        \dualq{b}{i} = \q{b}{i} + \dualunit \frac{1}{2} \qpos{b}{i}{i} \q{b}{i}, \quad \dualo{b}{i}{i} = \qomega{b}{i}{i} + \dualunit (\qvel{b}{i}{i} + \qomega{b}{i}{i} \cross \qpos{i}{b}{i}) \\
        \dualq{d}{i} = \q{d}{i} + \dualunit \frac{1}{2} \qpos{d}{i}{i} \q{d}{i}, \quad \dualo{d}{i}{i} = \qomega{d}{i}{i} + \dualunit (\qvel{d}{i}{i} + \qomega{d}{i}{i} \cross \qpos{i}{d}{i})
    \end{gathered}
\end{equation}
Then the initial relative dual state is then calculated by
\begin{equation}
    \begin{gathered}
        \dualq{b}{d} = \qconj{\dualq{d}{i}} \dualq{b}{i} \\
        \dualo{b}{i}{b} = \qconj{\dualq{b}{i}} \dualo{b}{i}{i} \dualq{b}{i}, \quad \dualo{d}{i}{b} = \qconj{\dualq{b}{i}} \dualo{d}{i}{i} \dualq{b}{i} \\
        \dualo{b}{d}{d} = \dualo{b}{i}{b} - \dualo{d}{i}{b}
    \end{gathered}
\end{equation}

The state of the desired relative to inertial is numerically integrated in parallel with the relative motion. The equations of motion for the desired relative to inertial are:
\begin{align}
    \ddualq{d}{i} &= \frac{1}{2} \dualo{d}{i}{i} \dualq{d}{i}  \\
    \ddualo{d}{i}{i} &= \qalpha{d}{i}{i} + \dualunit \left(\qacc{d}{i}{i} - \qalpha{d}{i}{i} \cross \qpos{d}{i}{i} - \qomega{d}{i}{i} \cross \qvel{d}{i}{i} \right)
\end{align}
The dynamics are derived using the dual quaternion version of the transport theorem and the angular acceleration is computed analytically from \cite{Filipe2015}
\begin{align}
    \qalpha{d}{i}{i} = \dqomega{d}{i}{i} = \frac{(\qpos{d}{i}{i} \cross \qacc{d}{i}{i}) \norm{\qpos{d}{i}{i}}^2 - 2 (\qpos{d}{i}{i} \cross \qvel{d}{i}{i}) (\qpos{d}{i}{i} \qdot \qvel{d}{i}{i})}{\norm{\qpos{d}{i}{i}}^4}
\end{align}
The desired spacecraft experiences the same external forces as the target, so the external accelerations acting on the desired are:
\begin{align}
    \qacc{d}{i}{i} = \qaccdes{g} + \qaccdes{\jtwo}
\end{align}

\subsection{Gramian Calculation}
The state vector used for the empirical observability Gramian calculations should consist of all independent states we wish to perform an observability analysis on and possibly estimate at a later time. For the relative proximity operation problem, we wish to investigate the pose and velocities of the body relative to the desired. This information is embedded in the relative dual position and velocities and the state vector for the empirical Gramian calculation is:
\begin{align}
    \label{eq:state}
    x = \mtx{\dualq{b}{d} \\ \dualo{b}{d}{b}} \in \reals{16}
\end{align}
However, the first entries of the real and dual components of the dual velocity quaternion will always be 0, so there are actually only 14 independent states in the state vector that are perturbed. If $x$ is thought of as a stacked 16-dimensional vector, then entries 9 and 13 correspond to the 0 components.

The empirical observability Gramian algorithm returns a Gramian matrix for each measurement in the system. In this scenario, there are 54 candidate tags and each tag returns a measurement in $\reals{5}$; meaning there are a total of 270 Gramian matrices calculated. The five matrices corresponding to the respective range and relative quaternion measurement of each tag are summed together to reflect the cumulative amount of information stored in a single tag; reducing the total number of Gramians to 54 and matching the intuition of one Gramian per tag. Lastly, when a tag does not meet the minimum elevation constraint at a measurement sample time it is equivalent to summing with a zero matrix for the empirical Gramian calculation at that time step.

\subsection{Results}
\begin{figure}[ht]
    \centering
    \includegraphics[height=3in]{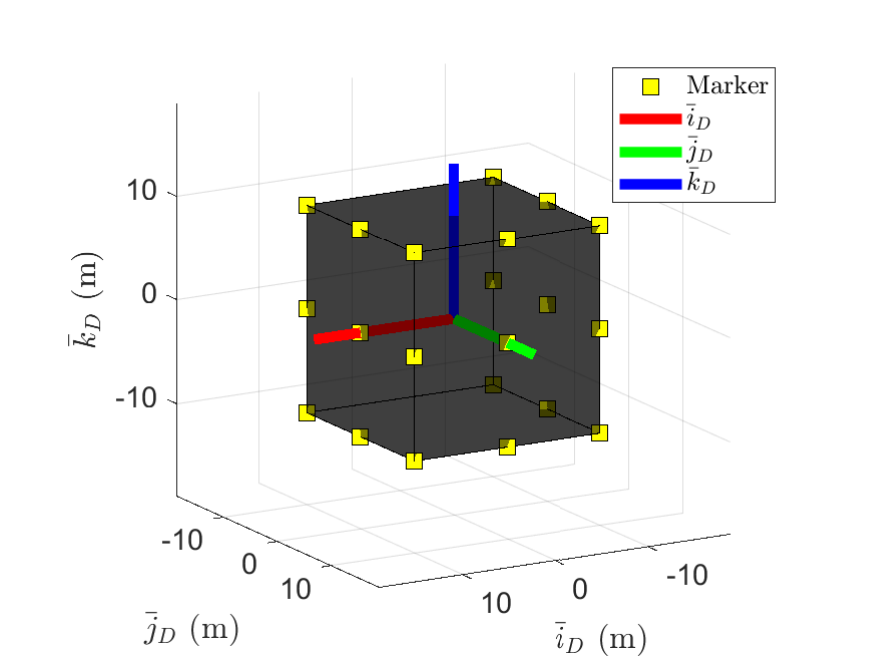}
    \caption{Visible fiducial marker set on the desired spacecraft.}
    \label{fig:vistag}
\end{figure}

\begin{figure}[ht]
    \centering
    \makebox[\textwidth][c]{
    \includegraphics[height=3in]{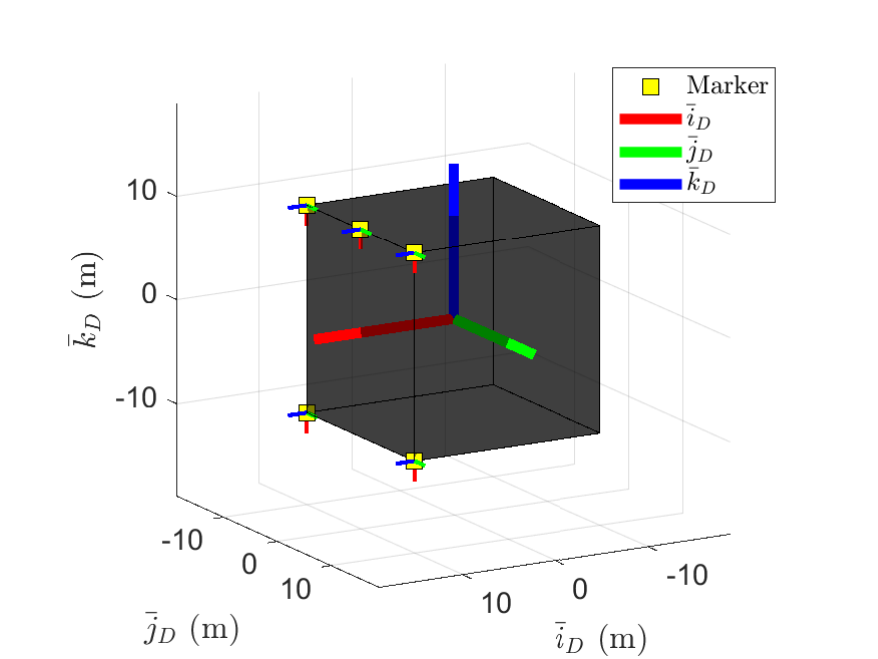}
    \includegraphics[height=3in]{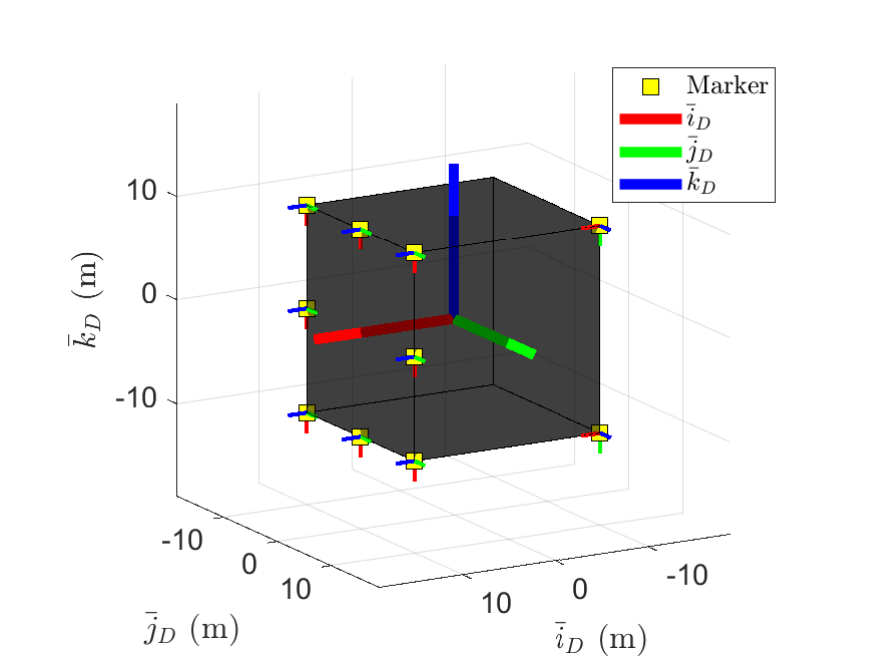}
    }
    \caption{Optimal fiducial marker set of five (left) and ten (right) on the desired spacecraft.}
    \label{fig:opttag}
    \makebox[\textwidth][c]{
    \includegraphics[height=3in]{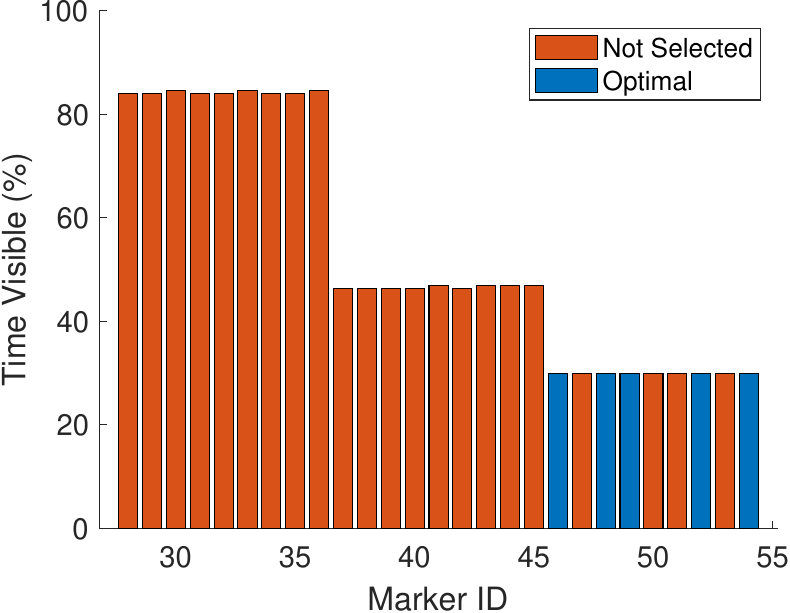}
    \includegraphics[height=3in]{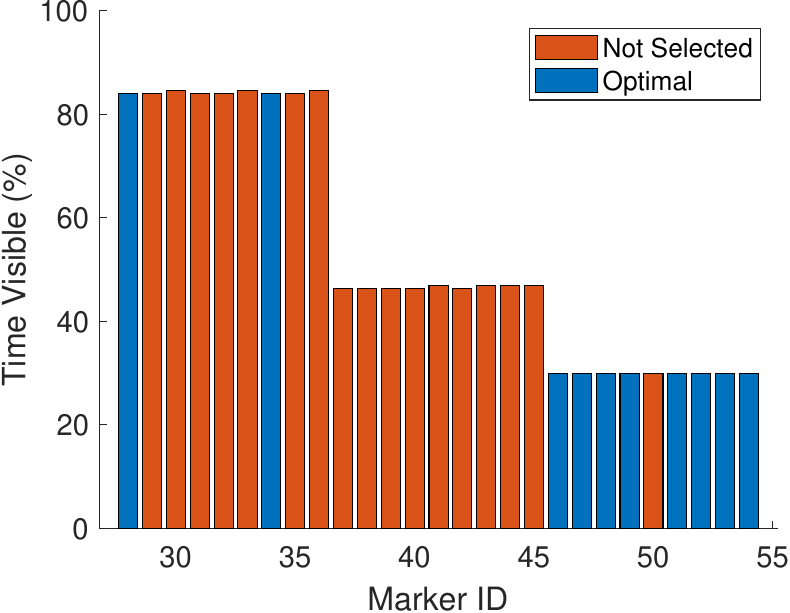}
    }
    \caption{Percent of time visible for optimal fiducial marker set of five (left) and ten (right).}
    \label{fig:vistime}
\end{figure}

Of the 54 tags in the candidate set, only 27 are visible at least once during the flyby. The visible set lies on the $+\cfvec{i}{d}$, $+\cfvec{j}{d}$, and $-\cfvec{i}{d}$ planes and is shown in figure \ref{fig:vistag}. The empirical observability Gramian optimal sensor placement problem was solved for a set of five and ten fiducial markers; results are shown in figure \ref{fig:opttag}. In both cases, the solution prefers to place tags at corners and allocates most of the tags to the $+\cfvec{i}{d}$ face. Placing tags at the corners maximizes the distance between tags, thus giving the greatest diversity in range and relative quaternion measurements, and therefore increases the ability to resolve a unique solution for the relative state. 

Alternative approaches to tag placement could be to place tags evenly across all visible faces or to place tags weighted by how often they are visible. An interesting result from using the observability Gramian for optimal tag placement is that the $+\cfvec{i}{d}$ face is favored over the other two visible faces, despite it being visible for the least amount of time between the three. The $+\cfvec{i}{d}$ face is only visible during the flyby approach, but the nonlinear dynamics during this phase are very sensitive to changes in the initial state, and maximizing observability is synonymous with selecting sensors that are most sensitive to measuring changes in state. Despite there being fewer measurements taken off the $+\cfvec{i}{d}$ face, these measurements provide more valuable information when trying to resolve a unique initial state from a set of measurements. This counterintuitive result demonstrates the power of an observability-driven optimal sensor placement methodology. A comparison of visibility time versus selected tags is shown in figure \ref{fig:vistime}.

\section{Conclusion}
This paper applied the observability-based optimal sensor placement problem to a dual quaternion satellite relative motion model. First, background material on quaternions, dual quaternions, and observability Gramians was introduced. Optimal fiducial marker locations on the surface of the observed satellite during a geostationary flyby were then solved for sets of five and ten markers. In both sets, it was shown that the optimal solution prefers to maximize the distance between markers. Contrary to intuition, the optimal solution does not necessarily align with the candidate sensor locations that are visible for the most amount of time. Maximizing the observability of the system through optimal sensor selection is synonymous with selecting the set of sensors that are most sensitive to measuring changes in state, which will ultimately minimize the estimate covariance.

Future work will test the performance of the optimal sensor set versus the suboptimal set in simulation and on a physical system. In both cases, measurements will be taken with both sets, processed through a Kalman filter, and the estimated error covariance will be compared. As a part of this work, some time will be spent characterizing the performance of fiducial markers when affected by real-world visibility constraints and noise such as shadowing, surface curvature, and tag pattern design. Lastly, there are some challenges that arise when calculating the empirical observability Gramian over long-duration trajectories. Gramian scaling methods to handle these long trajectories and better condition the semidefinite program will be investigated.

\section{Acknowledgment}
The authors would like to thank Nathan Powel for technical guidance and Panagiotis Tsiotras for providing his MATLAB dual quaternion toolbox. Contact the primary author at \href{mailto:nian6018@uw.edu}{nian6018@uw.edu} for MATLAB code used to generate numerical results. This research was partially funded by Blue Origin. 

\clearpage
\section{Notation}
\begin{tabular}{r l}
    $\qvec{\cdot}$ & vector in $\reals{3}$ \\
    $\cf{a}$ & coordinate frame \\
    $\body$ & body satellite coordinate frame \\
    $\desired$ & desired satellite coordinate frame \\
    $\inertial$ & Earth-centered inertial coordinate frame \\
    $\eye{n}$ & $\reals{n \times n}$ identity matrix \\
    $\mu$ & gravitational parameter \\
    $\jtwo$ & Earth oblateness constant \\
    $\inertia$ & $\reals{3 \times 3}$ body frame inertia matrix \\
    $\mass$ & body frame dual inertia matrix \\
    $\dualunit$ & dual unit \\
    $\qzeros$ & zero quaternion  $(0, \ \zeros{3}{1})$ \\
    $\qones$ & one quaternion  $(1, \ \zeros{3}{1})$ \\
    $\dualzeros$ & zero dual quaternion $\qzeros + \dualunit \qzeros$ \\
    $\dualones$ & one dual quaternion $\qones + \dualunit \qzeros$ \\
    $\pos{x}{y}{z}$ & position vector of $\cf{x}$ relative to $\cf{y}$ in $\cf{z}$ frame coordinates frame coordinates \\
    $\qpos{x}{y}{z}$ & position quaternion of $\cf{x}$ relative to $\cf{y}$ in $\cf{z}$ frame coordinates frame coordinates \\
    $\vel{x}{y}{z}$ & velocity vector of $\cf{x}$ relative to $\cf{y}$ in $\cf{z}$ frame coordinates \\
    $\qvel{x}{y}{z}$ & velocity quaternion of $\cf{x}$ relative to $\cf{y}$ in $\cf{z}$ frame coordinates \\
    $\q{x}{y}$ & unit quaternion of $\cf{x}$ relative to $\cf{y}$ \\
    $\dq{x}{y}$ & time derivative of unit quaternion of $\cf{x}$ relative to $\cf{y}$ \\
    $\dualq{x}{y}$ & unit dual quaternion of $\cf{x}$ relative to $\cf{y}$ \\
    $\ddualq{x}{y}$ & time derivative of unit dual quaternion of $\cf{x}$ relative to $\cf{y}$ \\
    $\omeg{x}{y}{z}$ & angular velocity vector of $\cf{x}$ relative to $\cf{y}$ in $\cf{z}$ frame coordinates \\
    $\qomega{x}{y}{z}$ & angular velocity quaternion of $\cf{x}$ relative to $\cf{y}$ in $\cf{z}$ frame coordinates \\
    $\dqomega{x}{y}{z}$ & time derivative of angular velocity vector of $\cf{x}$ relative to $\cf{y}$ in $\cf{z}$ frame coordinates \\
    $\dualo{x}{y}{z}$ & dual velocity of $\cf{x}$ relative to $\cf{y}$ in $\cf{z}$ frame coordinates \\
    $\ddualo{x}{y}{z}$ & time derivative of dual velocity of $\cf{x}$ relative to $\cf{y}$ in $\cf{z}$ frame coordinates \\
    $\acc{}$ & acceleration vector in $\cf{b}$ frame \\
    $\qaccbody{}$ & acceleration quaternion in $\cf{b}$ frame \\
    $\dualaccbody{}$ & dual acceleration vector in $\cf{b}$ frame \\
    $\force{}$ & force vector in $\cf{b}$ frame \\
    $\qforce{}$ & force quaternion in $\cf{b}$ frame \\
    $\dualforce{}$ & dual force in $\cf{b}$ frame \\
    $\torque{}$ & torque vector in $\cf{b}$ frame \\
    $\qtorque{}$ & torque quaternion in $\cf{b}$ frame \\
    $\og$ & analytical observability Gramian \\
    $\eog$ & empirical observability Gramian
\end{tabular}


\bibliographystyle{AAS_publication}   
\bibliography{references}   

\end{document}